# Measuring magnetic profiles at manganite surfaces with monolayer resolution

A. Verna[1], B. A. Davidson[1,*], Y. Szeto[2], A. Yu. Petrov[1], A. Mirone[3], A. Giglia[1], N. Mahne[1] and S. Nannarone[1,4]

[1] CNR-INFM TASC National Laboratory, Area Science Park (Basovizza), 34012 Trieste, Italy
[2] Dept. of Physics, Delft University of Technology, 2628 CJ Delft, The Netherlands
[3] European Synchrotron Radiation Facility, 6 rue Jules Horowitz, Boîte Postale 220, 38043 Grenoble, France
[4] Dipartimento di Ingegneria dei Materiali e dell'Ambiente, Università di Modena e Reggio Emilia, Via Vignolese 905, 41100 Modena, Italy

The performance of manganite-based magnetic tunnel junctions (MTJs) has suffered from reduced magnetization present at the junction interfaces that is ultimately responsible for the spin polarization of injected currents; this behavior has been attributed to a magnetic "dead layer" that typically extends a few unit cells into the manganite. X-ray magnetic scattering in resonant conditions (XRMS) is one of the most innovative and effective techniques to extract surface or interfacial magnetization profiles with subnanometer resolution, and has only recently been applied to oxide heterostructures. Here we present our approach to characterizing the surface and interfacial magnetization of such heterostructures using the XRMS technique, conducted at the BEAR beamline (Elettra synchrotron, Trieste). Measurements were carried out in specular reflectivity geometry, switching the left/right elliptical polarization of light as well the magnetization direction in the scattering plane. Spectra were collected across the Mn $L_{2,3}$ edge for at least four different grazing angles in order to better analyse the interference phenomena. The resulting reflectivity spectra have been carefully fit to obtain the magnetization profiles, minimizing the number of free parameters as much as possible. Optical constants of the samples (real and imaginary part of the refractive index) in the interested frequency range are obtained through absorption measurements in two magnetization states and subsequent Kramers-Kronig transformation, allowing quantitative fits of the magnetization profile at different temperatures. We apply this method to the study of air-exposed surfaces of epitaxial $La_{2/3}Sr_{1/3}MnO_3$ (001) films grown on $SrTiO_3$ (001) substrates.

*corresponding author: davidson@tasc.infm.it

## 1. Introduction

Transition-metal oxides are promising materials for spin electronic devices, some displaying half-metallic behavior [1] that make them ideal as sources and collectors of spin-polarized currents, along with a high Curie temperature $T_C$. Perovskites based on manganese, or manganites, have been theoretically predicted to be half-metals, or more precisely, transport half-metals because of the localization of minority carriers near the Fermi level [2]. There is still controversy over the interpretation of the experimental data in this regard [3] that is of central importance to clarify the nature of ferromagnetism in these materials, and their potential for devices. The simple double-exchange (DE) model originally proposed for manganites [4], while capturing the essential physics underlying the simultaneous metal-insulator transition (MIT at the "peak" temperature $T_P$) and paramagnetic-ferromagnetic transition (at the Curie temperature $T_C$) in some of the manganite phases, must be modified to account for additional charge and lattice degrees of freedom that influence the ground state [5]. These competing mechanisms ultimately produce the variegated energy landscapes that give the manganite families one of the richest set of phase diagrams amongst strongly-correlated systems [6].

Magnetic devices typically depend on the properties of interfaces for their functionality; this is most evident in tunneling devices, in which the tunneling current depends on the interfacial density of states within 2-4 Å of the electrode/barrier interface. Thus, understanding the influence of interfaces on the weakening of ferromagnetic correlations is essential to improve the behavior of spintronic devices. Growth of epitaxial interfaces of manganites is facilitated by the wide variety of available insulating lattice-matched perovskites, including other manganite phases. All studies so far show weakened ferromagnetic correlations at these interfaces or surfaces as compared to bulk, except possibly at the lowest temperatures [3, 7]. This behavior, often ambiguously called the "dead layer" and by implication that the underlying mechanism is the same in all cases, has been interpreted in different scenarios as due to (lateral) nanoscale phase separation [8], DE in reduced dimensions [9], or redistribution of orbital occupancies (or "orbital reconstruction") [10], to mention a few of the most recent approaches in the literature. While adequate understanding of the behavior of the interfacial magnetization will likely come only by applying a wide range of complementary techniques, one aspect that was missing until recently was a method to measure the out-of-plane magnetization profile at these interfaces with monolayer resolution.

This capability was recently demonstrated by X-ray resonant magnetic scattering (XRMS) on the vacuum/surface interface of cleaved bilayer manganite single crystals [11]. In that study, comparison of the experimental reflectivity data was made qualitatively to simulated spectra using a postulated magnetization profile. The qualitative criteria used was the inversion angle for the dichroic RMS signal, that is, the incidence angle at which the dichroic response on the $L_3$ peak inverts its sign, from being mostly positive (smaller incidence angles) to mostly negative (larger incidence angles). This sign change in the dichroic response with scattering angle is due to the changing interference conditions between s-polarized and p-polarized components of the reflected light. For an infinitely thick sample, the angle at which the dichroic response inverts its sign can be roughly be interpreted as inversely proportional to the dead layer thickness, i.e., a thicker nonmagnetic layer should show inversion of the dichroic signal's sign at a smaller scattering angle, while a thinner dead layer should show inversion at a larger angle. This "inversion angle" can be more precisely defined as the angle at which the $L_3$ XRMS signal is half positive and half negative, i.e., its integral is zero. Below this angle, the $L_3$ integral is positive, while above this angle it is negative.

These considerations don't necessarily apply to XRMS spectra of thin film heterostructures, which contain additional effects that can strongly influence the resonant reflectivity and yield an XRMS response far different from the simple model suggested above for a infinite crystal. These effects can result from scattering from multiple interfaces, thickness effects of the different layers, chemical variations along the surface normal, and buried dead layers, to name a few. Accurate information on the magnetization profile can only be obtained by fitting the curve over the entire energy range, including both $L_3$ and $L_2$ peaks, for several scattering angles, as discussed below.

After [11], XRMS studies were performed on manganite thin films, reporting fits of the experimental data measured at one scattering angle to obtain the magnetization profile [12], [13]. Two different structures were investigated: uncapped LCMO (C=Ca, bulk $T_C$ ~ 200-250 K) [13] exposed to air for at least some days; and $La_{2/3}Sr_{1/3}MnO_3$ (LSMO, bulk $T_C$ ~ 360 K) with an epitaxial $SrTiO_3$ (STO) cap layer [12]. The air-exposed LCMO samples showed a thick nonmagnetic layer (~1-2nm) at the surface at 80 K. Surprisingly, the capped LSMO samples showed a weakened but still magnetic layer up to the interface at 300 K (~0.15 of the bulk magnetization $M_{bulk}$ at the interface), and an atomically-engineered underdoped interface showed larger magnetization than the "as-grown" interface but only at low tempertures (~0.7 $M_{bulk}$ at 50 K versus 0.5 $M_{bulk}$).

Here we present our approach to extracting magnetization profiles from XRMS data, utilizing a single (minimal) set of parameters to fit spectra taken at four different scattering angles. We studied uncapped LSMO samples exposed to air for several weeks or months between the growth and XRMS measurements. Several parameters were independently measured, such as surface roughness and film thickness, thereby reducing the number of free parameters in the fit.

## 2. Experimental

The $La_{1-x}Sr_xMnO_3$ (x=0.33) films studied here were grown by ozone-assisted molecular beam epitaxy (MBE) at 740°C on $TiO_2$-terminated STO (001) substrates in a partial pressure of $8 \times 10^{-7}$ mBar pure ozone. Absolute calibration of the fluxes to the level of a few percent was performed by Rutherford Backscattering Spectrometry (RBS) and periodically rechecked by RBS and/or by fitting finite thickness oscillations measured by glancing incidence X-ray diffraction (GIXRD). For the data in Figs. 1 and 2, GIXRD gave a film thickness of 366 Å (within 1.5% of that expected from the RBS calibration). Atomic flatness was optimized *in situ* by reflection high-energy electron diffraction analysis, and quantified *ex situ* by atomic force microscopy. RMS surface roughness was ~1 Å over 2x2 µm² scans, determined by the step-terrace structure of the STO substrate. The film showed metallic behavior up to at least 500K, and the maximum derivative of the resistivity versus temperature occurred near ~345 K.

The XRMS measurements were performed at the BEAR beamline of Elettra Synchrotron in Trieste. A special home-made sample holder was built that permitted application of an in-plane magnetic field up to ~600 Oe on the sample during measurements. The saturation field of the films is typically less than 600 Oe and the hysteresis loops are very nearly square. With liquid nitrogen cooling, temperatures of 200 K at the sample were reached. The degree of circular polarization of the light was measured using Stokes polarimetry at ~90%, selecting the appropriate part of the beam with polarization slits after the bending magnet. Careful sample alignment ensured accuracies of ~0.1° in the scattering angle.

XRMS spectra were acquired in specular reflection geometry at various grazing angles with respect to the sample surface (here in the range 7°-15°), measuring the intensity of the reflected radiation as a function of the photon energy across the Mn $L_{2,3}$ absorption edge. Absolute reflectivity spectra are given by the ratio between the photodiode yield in reflection and the yield obtained when the X-ray beam directly impinges on the photodiode, both normalized to the drain current of a gold mesh intersecting the beam before the sample that is proportional to the photon flux. For each grazing angle, two reflectivity scans are measured in remnance using right-hand circularly polarized light, after having magnetized the sample in the two opposite directions in the scattering plane. XRMS signal is simply given by the difference of the reflectivities for the two magnetization states ($r^+ - r^-$), without any further manipulation of the experimental data.

The reflectivity spectra and XRMS signals are fitted using the Pythonic Programming for Multilayer code [14] that it is based on an extension of the Parratt formalism [15] to optically anisotropic materials. The optical properties of the magnetic LSMO epilayer across the Mn $L_{2,3}$ absorption edge are characterized by two different complex refractive indexes, $N^+(\omega)=n^+(\omega)+i\beta^+(\omega)$ and $N^-(\omega)=n^-(\omega)+i\beta^-(\omega)$, relative to two opposite orientations of the magnetization with respect to the photon helicity. The difference $\Delta N = N^+ - N^-$ between the two refractive indexes is directly proportional to the magnetization, and, due to the presence of the dead layer, is reduced near the surface, while the average $\bar{N} = 1/2(N^+ + N^-)$ depends only on the chemical properties of the LSMO film. The dichroic Mn contribution to the imaginary parts $\beta^+$ and $\beta^-$ is inferred via X-ray absorption spectroscopy (XAS) measurements, performed in total electron yield (TEY) in the two remnant magnetization states. After a small correction for saturation effects [16], absorption spectra in TEY are directly proportional to the absorption coefficients $\mu^\pm = 4\pi\beta^\pm/\lambda$, with the proportionality constant treated as a fitting parameter. The Mn contribution to the real part $n^\pm$ of the refractive index is then obtained through the Kramers-Krönig relations. For the contribution of the other elements (La, Sr, O, Ti), which have absorption edges far from the energy region of interest, tabulated values are used [17]. It is important to note that, for our system, this method underestimates the dichroic effect $\Delta N$ present in the bulk of the LSMO layer, because in TEY measurements electrons are also extracted from the dead layer, but the energy dependence of $\Delta N$ is correctly reproduced. For the fitting procedure, the near-surface region of the LSMO film was divided in layers with the thickness of 1 unit cell (out-of-plane lattice parameter c=3.83 Å as measured by XRD) having the same chemical composition but different magnetization values (and thus different dichroism in the refractive index). Reflectivity spectra were then fitted by varying the magnetization of the surface monolayers and of the bulk, obtaining the correct magnetization profile of the sample. In order to minimize the number of free parameters in the fitting, the total film thickness and the RMS surface roughness were measured independently as previously described, and the (buried) film/substrate interface roughness was set to 1 Å.

### 3. Results and discussion

We measured XRMS spectra on several air-exposed, uncapped LSMO films and present representative results in Fig. 1. The experimental data were taken at 7°, 9°, 12° and 15° incidence angles and at two different temperatures, 200 and 300 K, from a film whose thickness is 366 Å. The magnetization profiles that produce the fits to the data (Fig. 1, solid lines) are shown in Fig. 2 for both temperatures, revealing a fully "dead layer" of ~17 Å at 200 K (and a profile width 10%-90% of ~3 unit cells) and ~21 Å at 300 K. In the fits to the data that generate the profiles shown in Fig. 2, the only free parameters are the magnetization in each of the first 10 unit cells of LSMO, the bulk magnetization, and the proportionality constant (always close to 1) that relates the absorption coefficient $\mu$ to the XAS data, as discussed above. We note that a *single set of parameters* for these values produced the fits at different angles shown in Fig.1 for each temperature, in which the absolute value of the reflectivity changes by more than a factor of 20 ($I \propto \sin^{-4}(\Theta)$, where $\Theta$ is the scattering angle).

The profile extracted from the data at 300 K is obtained independently of the profile at 200 K, but nevertheless the profiles have similar shape, showing a roughly one-unit-cell shift to larger "dead layer" thickness, and a reduced saturation magnetization as temperature is increased. Both of these trends are physically reasonable. As can be seen from Fig. 1, the thicker "dead layer" at 300 K is also seen in the experimental data of Fig. 1 as a shift of the inversion angle to lower angle, i.e., the angle at which the XRMS $L_3$ signal is roughly half positive and half negative is ~7° at 200 K and clearly <7° at 300 K.

It can be noticed in Fig. 1 that fits at higher angles, e.g., 15º, show somewhat larger error than smaller angles. This could be due in part to the fact that no "dead layer" was assumed at the film/substrate interface, as might be expected to exist. Including as a fitting parameter the thickness of such a buried "dead layer" can improve the agreement with the experimental data but this is only apparent at higher angles for such thick films, and the improvement is small. Since the absorption length for 640 eV photons is ~100 nm in these materials, contributions to the reflection intensity from the buried interface are negligible at smaller incidence angles.

The film shown in Fig. 1 was exposed to air for several months before the XRMS measurements were performed; characterization of a similar film measured after only a couple weeks of air exposure showed a thinner "dead layer" of ~12 Å at 200 K. Further studies need to be performed in order to understand the correlation of air exposure to the "dead layer" thickness.

To demonstrate that the XRMS fits are quite sensitive to the nonmagnetic layer thickness, in Fig. 3 we show the experimental data (at 200 K and 7°) and the simulated data that result when the best-fit profile from Fig. 2 (blue curve) is shifted one unit cell towards or away from the surface. The XRMS curves change significantly as the profile is shifted in either direction, and the disagreement with the experimental data is quite apparent. This allows us to conclude that our approach to identify the magnetization profile has a spatial resolution much better than 4 Å. Again, the relationship between inversion angle and "dead layer" thickness is evident, since the inversion angle in the dashed-line XRMS spectra has obviously moved to higher angle, and vice versa for the dash-dot-line spectra. Since the fitting is most sensitive around the inversion angle, the most reliable results are found when the experimental spectra are taken at enough angles to span the inversion angle (i.e., from below to above this angle).

The profiles shown in Fig. 2 are in reasonable agreement with recent XRMS results published on air-exposed LCMO films [13]. In those studies, seven fitting parameters were used. A step-function was assumed for the magnetization profile and the thickness of the fully "dead layer" was used as a fitting parameter, along with three roughness parameters (surface, film/substrate, and "dead layer"/magnetized layer roughness), film thickness, and two scaling factors for the absorption data for each region [13]. The XRMS signal was measured at 80 K and at 13°, close to the inversion angle for the ~15-20 Å "dead layers" in those films. The resulting surface nonmagnetic layer ranged from 5-21 Å, depending on the substrate, with fit RMS roughnesses of 1-29 Å at the surface or different buried interfaces, again depending on substrate.

This is in strong contrast to XRMS results on capped LSMO samples that never showed a fully "dead layer" up to 300 K, only a reduced but finite magnetization at the interface [12]. The data of [12] was taken at a single angle (11°) and it is notable that their reported XRMS spectrum (shown in Fig. 4 of [12]) is comparable to our data of Fig. 1(a) taken at 7°, showing a higher inversion angle and therefore a thinner "dead layer" in their capped film, as verified by the fit magnetization profiles. In [12] there is no discussion of the fitting parameters used to simulate the data, but the results clearly demonstrate the high quality of the heterostructures investigated, and that capped surfaces appear to be qualitatively different than free surfaces, as has been commented recently [18].

It should be noted that the magnetization measured experimentally here is a lateral average over the spot size of the synchrotron light beam (in our case, ~50x500μm$^2$), since nonferromagnetic regions don't contribute to the signal. The simulations of the reflectivity also assume a laterally homogeneous magnetization that varies only along the surface normal direction. The quite good fits to the data at 200 K suggest that such an assumption is valid at lower temperatures, while the more noticeable deviation of the fits at 300 K may be in part due to such lateral nanoscale phase separation, which are expected to be more important closer to $T_C$. When the phase separation is not too sharp and doesn't extend throughout the "dead layer" in a columnar-type structure, then including a parameter for the roughness between the magnetic and nonmagnetic layers may improve the fits.

## 4. Conclusions

The XRMS technique has been used to extract the magnetization profiles at air-exposed LSMO surfaces, showing that a fully magnetically "dead layer" exists at the surface. The thickness of this "dead layer" increases with temperature, growing from ~17 Å at 200 K to ~21 Å at 300 K, and the bulk magnetization is reduced as $T_C$ is approached. The form of the profile is shown to be relatively constant in temperature, shifting away from the surface as temperature is increased. Moreover, the effectiveness of the technique in estimating the "dead layer" thickness with better than one unit cell resolution has been demonstrated. This technique can be readily applied to heterostructures. In our modeling of the anisotropic optical constants, we used only an (average) gradient of the magnetization in the surface-normal direction, and so no lateral information could be recovered (viz. nanoscale phase separation into ferromagnetic and nonferromagnetic regions). Indeed, while different arguments have been offered for "dead layer" formation at the surfaces of manganites, XRMS results cannot offer explicit clues to the underlying mechanisms other than the (average) magnetization profile (shape and depth) along the surface-normal direction as it evolves with temperature. This by itself is an essential piece of information for any theoretical approach to describing the "dead layer". The presence of antiferromagnetic domains at the interface that grow as temperature is increased, a likely scenario, might be investigated in a complementary way by magnetic linear dichroism (XMLD) studies provided that other mechanisms that contribute to the LD signal can be separately identified and substracted.

## 5. Acknowlegements

The RBS measurements were done at the AN2000 and CN accelerators of the INFN Legnaro National Laboratories (Padova, Italy), and we thank Prof. M. Berti for the use of their backscattering goniometer and electronics. This work was supported by the FVG Regional project SPINOX funded by the Legge Regionale 26/2005.

**Figure 1.** XRMS signal, r$^+$ - r$^-$, as a function of photon energy across the Mn L$_{3,2}$ edge, where r$^+$ and r$^-$ indicate the reflectivity with the sample magnetized parallel or antiparallel to the photon helicity. a) measured at 200 K; 7°, 9°, 12° and 15° scattering angles; b) at 300 K; same angles. Open circles are the experimental data, red lines are the fits from PMM simulations for different angles. For each temperature, a single set of parameters is used for the fitting the spectra at all four angles (see discussion in the text).

**Figure 2.** Magnetization profiles that yield the fits shown in Fig. 1. Blue lines are for the data at 200 K, red lines for the data at 300 K. Since the simulation is divided into unit-cell layers, the proper profile is given by the steps; the curved lines are guides to the eye.

**Figure 3.** Demonstration of spatial resolution of the magnetization profiles from the fitting procedure. a) XRMS spectra corresponding to different magnetization profiles shown in b). The open circles are the experimental data at 7° and 200K, the solid line in a) is the fit XRMS spectrum generated by the profile shown as the solid line of b), reproduced from Figs. 1 and 2. The dashed-line XRMS spectrum in a) results from the dashed-line profile of b), which is the same profile but moved one unit cell (~4 Å) towards the surface. The dash-dot XRMS spectrum in a) results from the dash-dot profile in b), which is again the same profile as the solid line but moved one unit cell away from the surface.

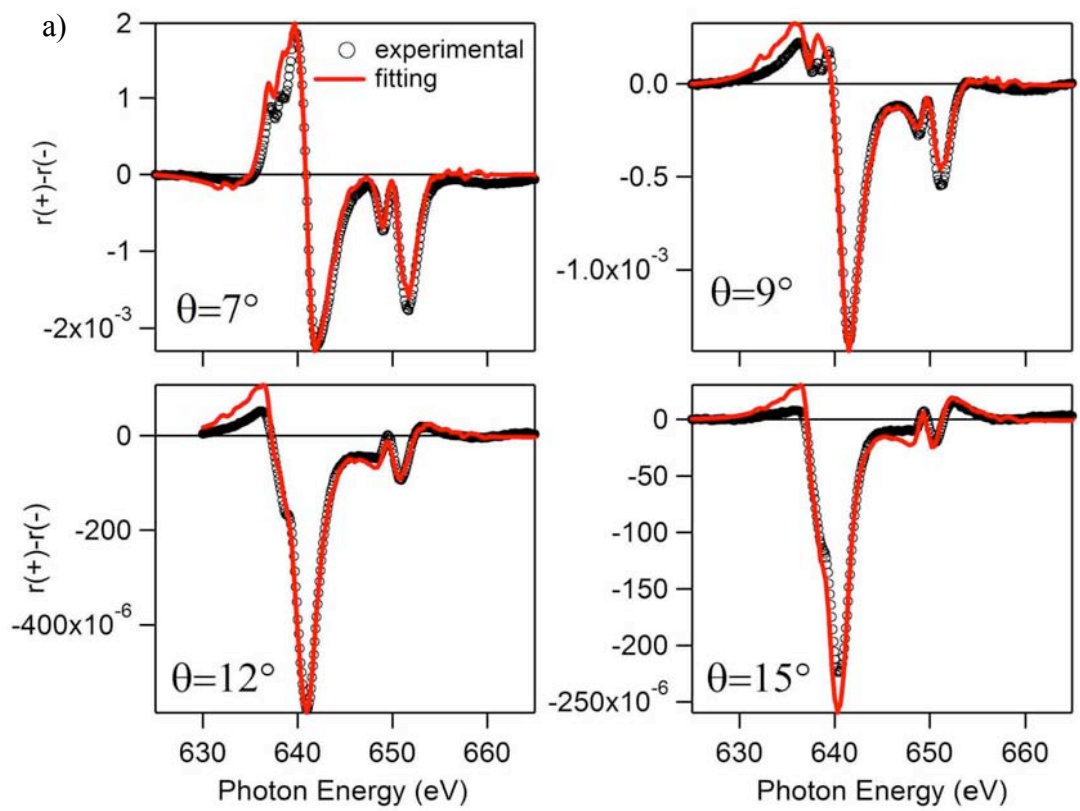

FIGURE 1(a)  Verna/Davidson

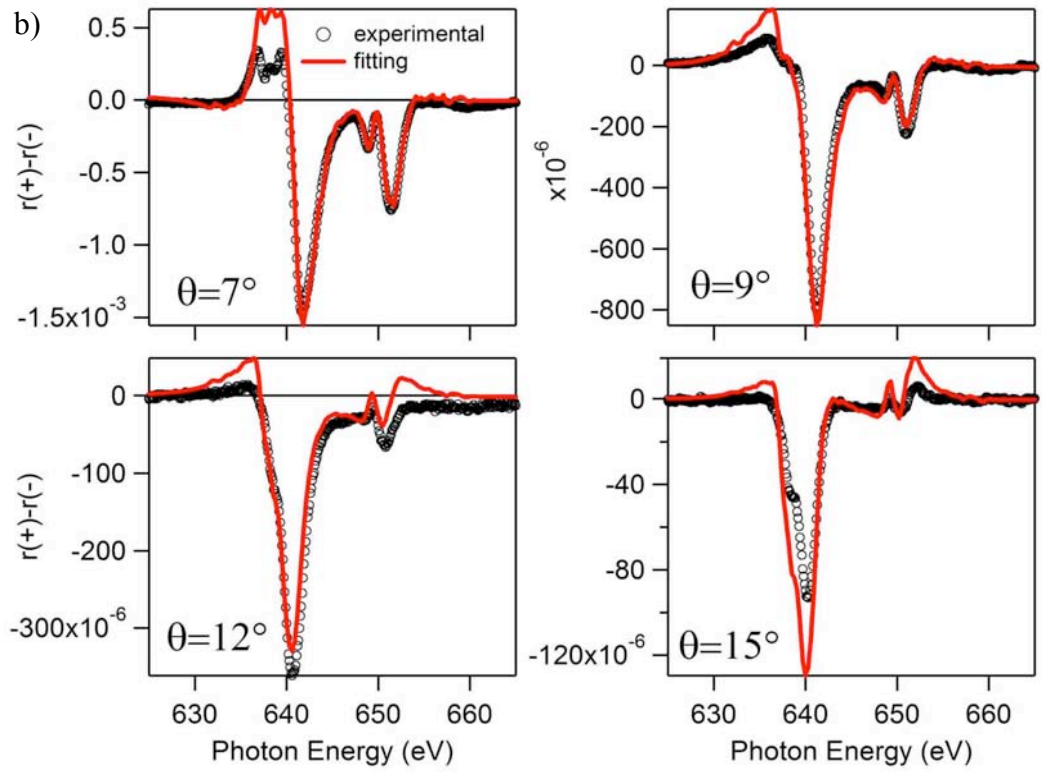

FIGURE 1(b) Verna/Davidson

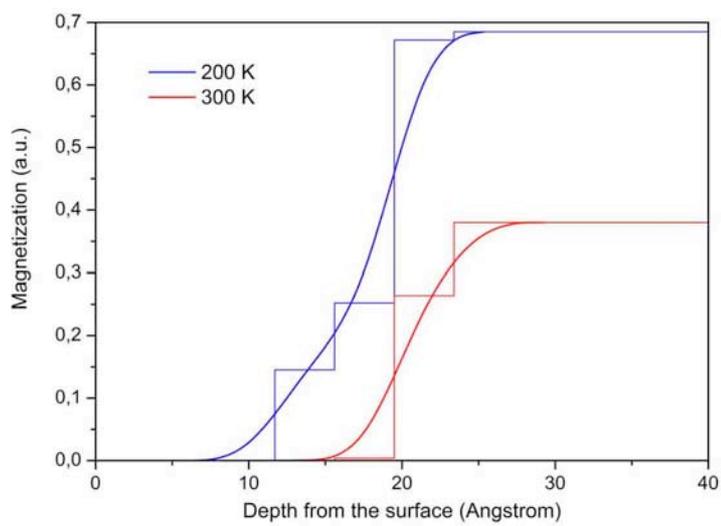

FIGURE 2                    Verna/Davidson

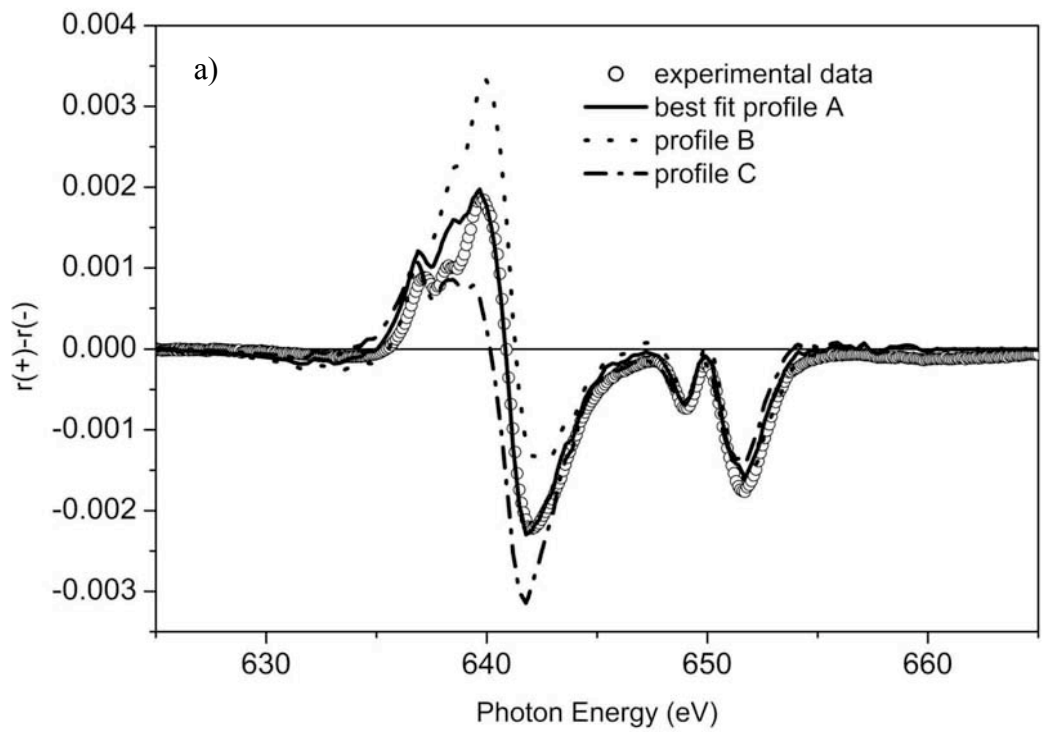

FIGURE 3(a)   Verna/Davidson

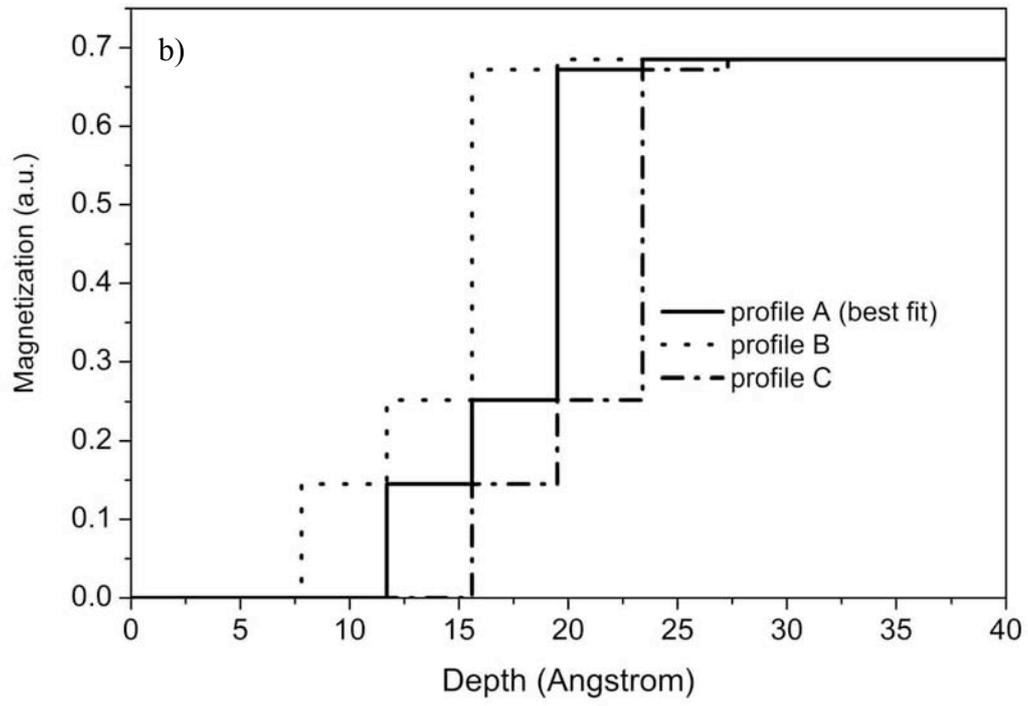

FIGURE 3(b)　　　　Verna/Davidson